\documentclass{article}

\usepackage{arxiv}

\usepackage[utf8]{inputenc} 
\usepackage[T1]{fontenc}    
\usepackage{hyperref}       
\usepackage{url}            
\usepackage{booktabs}       
\usepackage{amsfonts}       
\usepackage{nicefrac}       
\usepackage{microtype}      
\usepackage{lipsum}		
\usepackage{graphicx}
\usepackage{natbib}
\usepackage{doi}
\usepackage{caption}
\usepackage{subcaption}
\usepackage{algorithm}
\usepackage{algorithmic}

\title{Hyper-Realist Rendering: A Theoretical Framework}

\author{  \href{https://orcid.org/0000-0003-3618-4166}{\includegraphics[scale=0.06]{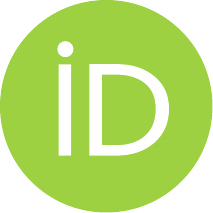}\hspace{1mm}Ergun Akleman}\thanks{Joint with Computer Science and Engineering Department.} \\
	Visual Computing \& Computational Media, School of PVFA\\
 Computer Science and Engineering, College of Engineering\\
 Texas A\&M University, College Station, TX, 77831\\
	\texttt{ergun@tamu.edu} \\
 	\And
	 \href{https://orcid.org/0000-0002-3236-5595}{\includegraphics[scale=0.06]{orcid.pdf}\hspace{1mm}Murat Kurt}\\
	 International Computer Institute \\ Ege University \\Izmir, Türkiye\\	\texttt{murat.kurt@ege.edu.tr} \\
  	\And
 \href{https://orcid.org/0009-0000-4712-1385}{\includegraphics[scale=0.06]{orcid.pdf}\hspace{1mm}Derya Akleman}\\
Department of Statistics\\
	Texas A\&M University, College Station, TX, 77831\\
	\texttt{akleman@tamu.com} \\
 \And
 \href{https://orcid.org/0009-0008-8109-2686}{\includegraphics[scale=0.06]{orcid.pdf}\hspace{1mm}Gary Bruins}\\
Visualization Program, Department of Architecture,\\
	Texas A\&M University, College Station, TX, 77831\\
	\texttt{garybruins@hotmail.com} \\
  \And
   	Sitong Deng\\
Department of Visualization\\ 
Texas A\&M University, College Station, TX, 77831\\
	\texttt{sitongdeng@tamu.edu} \\
    \And
 Meena Subramanian\\
Department of Visualization\\ 
	Texas A\&M University, College Station, TX, 77831\\
	\texttt{meena.subramanian18@tamu.edu} \\
}

\renewcommand{\shorttitle}{Hyper-Realist Rendering}

\hypersetup{
pdftitle={A template for the arxiv style},
pdfsubject={q-bio.NC, q-bio.QM},
pdfauthor={David S.~Hippocampus, Elias D.~Striatum},
pdfkeywords={First keyword, Second keyword, More},
}

\begin{document}
\maketitle
	
\begin{abstract}

This is the first paper in a series on hyper-realist rendering. In this paper, we introduce the concept of hyper-realist rendering and present a theoretical framework to obtain hyper-realist images. We are using the term Hyper-realism as an umbrella word that captures all types of visual artifacts that can evoke an impression of reality. The hyper-realist artifacts are visual representations that are not necessarily created by following logical and physical principles and can still be perceived as representations of reality. This idea stems from the principles of representational arts, which attain visually acceptable renderings of scenes without implementing strict physical laws of optics and materials. The objective of this work is to demonstrate that it is possible to obtain visually acceptable illusions of reality by employing such artistic approaches. With representational art methods, we can even obtain an alternate illusion of reality that looks more real even when it is not real. This paper demonstrates that it is common to create illusions of reality in visual arts with examples of paintings by representational artists. We propose an approach to obtain expressive local and global illuminations to obtain these stylistic illusions with a set of well-defined and formal methods. Our formalism is based on four principles: (1) colors and textures must be represented by algebraically complete structures, (2) we should allow impossible and inconsistent shapes in scene concepts, (3) we have to decompose illumination and shading, and (4) we must build our shading functions using statistic-based methods instead of physics-based methods. We claim that these principles allow us to develop simple methods since we can avoid algebraic inconsistencies. Using such simple methods, it can be possible to significantly speed up modeling and rendering algorithms by focusing only on essential elements that can create illusions of reality. 
		
\end{abstract}
	
\keywords{Expressive Rendering, Global Illumination, Computer Graphics}

\section{Introduction and Motivation}

The objective of Hyper-Realist Rendering is to study artistic approaches to identify the perceptual and cognitive requirements to obtain visually acceptable illusions of reality. Our idea stems from the principles of representational arts that attain visually acceptable renderings of scenes without implementing strict physical laws of optics and materials. With representational arts, we can even obtain an alternate illusion of reality that looks more real even when it is not real, which is usually called hyper-realism \cite{taylor2009, baudrillard1988, wolny2017}. Our goal is to formalize the methods of artists by turning them into a set of well-defined rules that provide the perceptual and cognitive requirements for hyper-realism. To reach this goal, one needs to test the validity of a set of hypotheses that we put forward based on the methods of representational artists. By formally identifying the requirements to obtain hyper-realistic results, it is possible to significantly speed up modeling and rendering algorithms for visualization by focusing only on essential elements that can create illusions of reality. 

According to Gombrich \cite{gombrich1960}, the processes of developing artistic methods demonstrate substantial similarity to the scientific processes, as presented by Popper's theory of science \cite{popper1995,richmond1994}. This similarity suggests that artistic methods can be used effectively to solve some difficult problems \cite{marsden1993,edens2007}. The advantage of artistic methods is that they provide simplified models that are particularly suitable for solving some specific problems, similar to scientific models. For vision-related problems, artistic methods are particularly appropriate, since representational painting, illustration, and sculpting all focus on the creation of an illusion of reality using intuitive and straightforward approaches.

The integral component of hyper-realist rendering will involve the formal identification of information that is essential for the creation of an illusion of reality. We already have a reasonable estimate of the information that is essential to capture based on the experience of representational artists. Our goal in this paper is to provide a general strategy for the development of specific hypotheses for any given artist or artistic style. These experience-based hypotheses and models will be developed using visual data-based statistical analysis. These hypotheses can also be evaluated with rigorous user studies to identify the quality of mathematıcal models.

 \begin{figure}
        \begin{subfigure}[t]{0.32\textwidth}
        \includegraphics[width=1.0\textwidth]{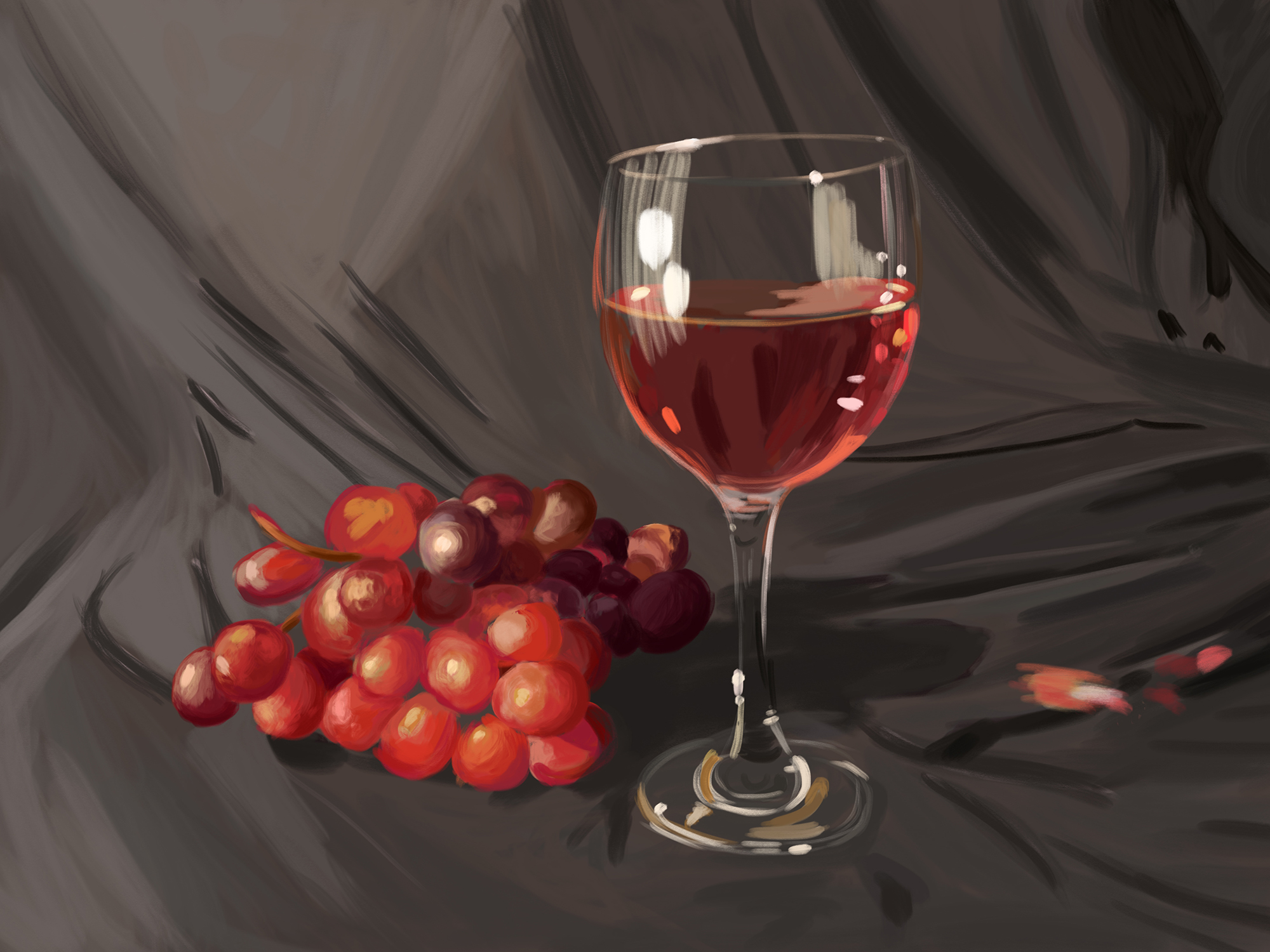}
        \caption{ Original Painting.}
        \label{fig_stillLifeOriginal}
    \end{subfigure}
    \hfill
         \begin{subfigure}[t]{0.32\textwidth}
        \includegraphics[width=1.0\textwidth]{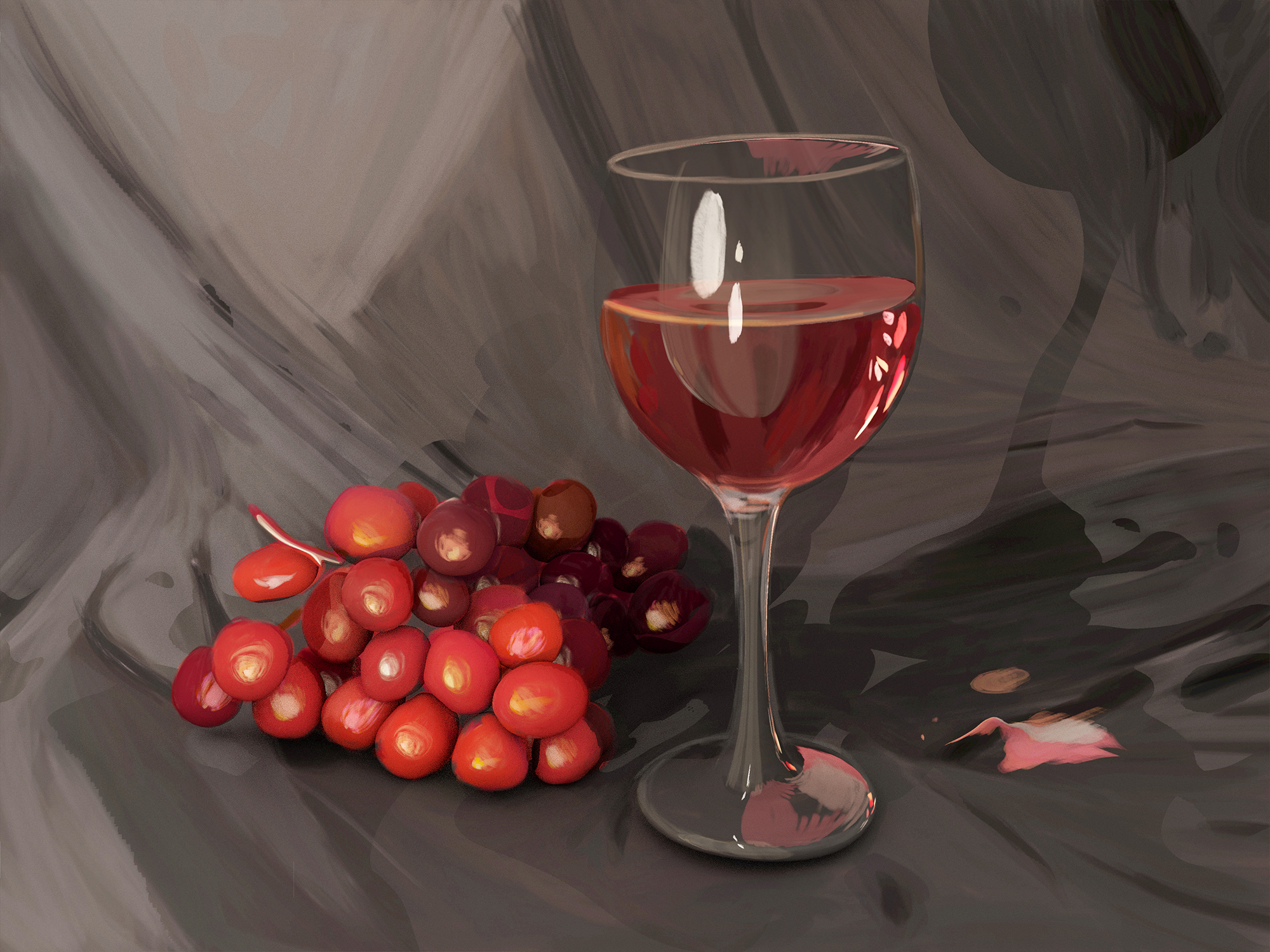}
        \caption{ Frame 15 of our animation.}
        \label{fig_frame 15}
    \end{subfigure}
    \hfill   
        \begin{subfigure}[t]{0.32\textwidth}
        \includegraphics[width=1.0\textwidth]{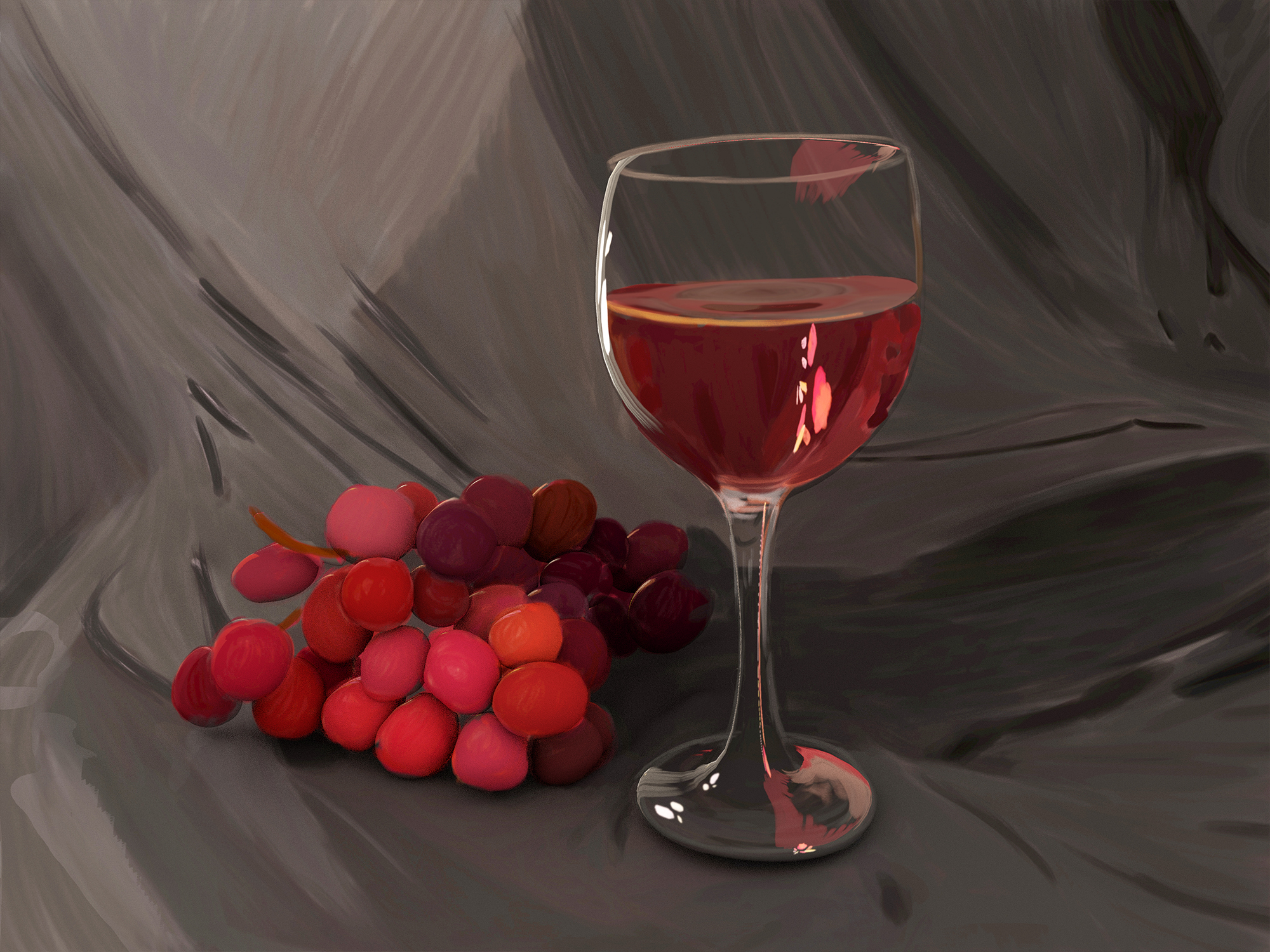}
        \caption{ Frame 110 of our animation.}
        \label{fig_frame110}
    \end{subfigure}
    \hfill
\caption{\it An example of hyper-realistic emulation: An animated still life painting with dynamically changing shadows, specular highlights, and caustics using a moving area light source that emulates a painting shown in Figure~\ref{fig_stillLifeOriginal}. The goal of such emulations is to provide direct artistic control for obtaining the desired look and feel. }
\label{fig_still_life}
\end{figure}

The global illumination effects are the most important elements in obtaining hyper-realism. The most well-known global illumination effects for hyper-realism are reflections, refractions, and caustics (see figures~\ref{fig_still_life} and \ref{fig_GaryBruins}). The main problem with these global effects is that obtaining these effects is also the hardest part of rendering computation. The difficulty comes from the fact that none of the global illumination effects can be created by using only local shader functions. Global illumination effects directly emerge from all the inter-reflections of lights in a given scene. To simulate global illumination effects, we should consider the relative positions and orientations of all objects (including air and water) in the given scene. The objects in a scene should usually include not only opaque solids, but also semi-transparent solids, liquids, and gases such as flesh, water, smoke, and even air. We also consider shadows as a part of global illumination effects, since they cannot be computed only by local shader functions. Similarly, scattering and subsurface scattering are also considered global effects, since all require more than local information to calculate the final image. In all of these cases, the interrelations between objects play an important role. 

\subsection{Basis and Rationale}
\label{BasisAndRationale}
 
Although the term hyper-realism is not commonly used, the goal of rendering in computer graphics practice has always been to obtain visual illusions of reality. The image generation process is well understood in current computer graphics practice: It consists of three main stages: (1)  Representation; (2) Rendering, and (3) Compositing \cite{hughes2014,kahrs1996,foley1994,brinkmann2008}. The representation stage involves the creation or reconstruction of (1) camera parameters \cite{hemayed2003}, (2) light positions and orientations \cite{debevec1998,debevec2002}, (3) orientations, positions and shapes of all objects in the scene \cite{zhang1999,slabaugh2001,hartley2003,stoykova2007}, and (4) material properties of all visible points \cite{filip2008, kurt2009survey, kurt2018DEU, kurt2019DEU}. 

With hyper-realism, our first conceptual contribution is to demonstrate that it is possible to significantly simplify the representations of objects and materials by using rough approximations that can provide the same illusions. An important implication of this simplification is that we can significantly speed up the representation and rendering processes. Our second conceptual contribution with the hyper-realistic rendering approach is to demonstrate that it is possible to split rendering into two parts, illumination and shading, by using algebraically consistent structures. An important implication of this split is that we can move shading to the compositing stage and directly control look-and-feel in post-processing, i.e. in the compositing stage. The hyper-realist approach can be used successfully in two of the key computer graphics applications: emulation and augmentation. 

\begin{figure}[ht]
    \centering
    \begin{subfigure}[t]{0.320\textwidth}
        \centering
        \includegraphics[width=1.0\textwidth]{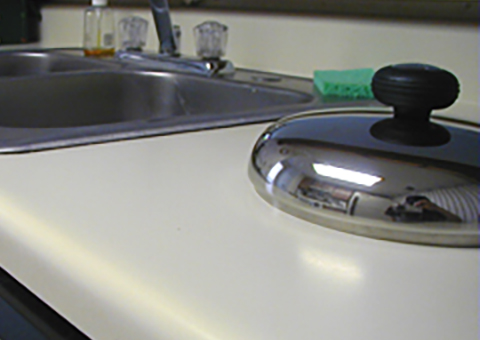}
         \caption{   A Photograph used as the Background.}
        \label{fig_GaryBruinsa}
    \end{subfigure}
    \hfill
        \begin{subfigure}[t]{0.320\textwidth}
        \centering
        \includegraphics[width=1.0\textwidth]{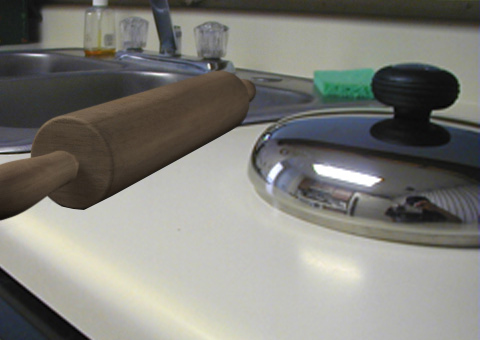}
         \caption{    A typical augmentation of the photograph with a virtual rolling pin.}
        \label{fig_GaryBruinsb}
    \end{subfigure}
   \hfill
    \begin{subfigure}[t]{0.320\textwidth}
        \centering
        \includegraphics[width=1.0\textwidth]{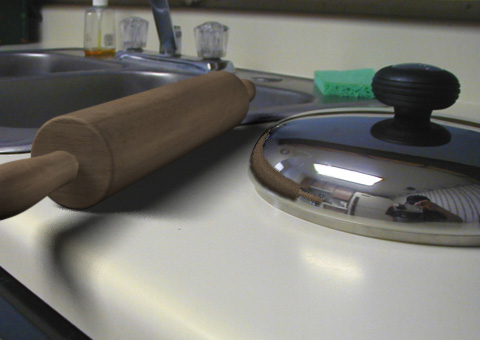}
         \caption{   A hyper-realistic augmentation of the photograph with a virtual rolling pin.}
        \label{fig_GaryBruinsc}
    \end{subfigure}
     \caption{   A hyper-realistic augmentation example that demonstrates demonstrates how global illumination improves overall quality. In this particular case,  the shadow and reflection of the rolling pin are added in \ref{fig_GaryBruinsc}. Reflection, in particular, becomes increasingly important during motion, such as moving the camera or object positions in AR applications. The augmented image in \ref{fig_GaryBruinsc} was created by Gary Bruins in a computer animation / digital compositing course in 2001 taught by Ergun Akleman. }
    \label{fig_GaryBruins}
\end{figure}

\begin{itemize}
\item \textbf{Hyper-Realistic Emulations:} These can be defined as a process of rendering hyper-realistic images by simulating scenes using virtual cameras, virtual lights, virtual geometry, and virtual materials that represent the scene. This is a standard computer graphics rendering process that includes all virtual reality applications. Figure~\ref{fig_still_life} demonstrates the importance of global illumination in emulation \cite{subramanian2020painterly}. The original still-life painting by Mena Subramanian contains a glass object, which can have reflection and refraction with Fresnel effect and grapes, which can have reflection and subsurface scattering. Therefore, an emulation must include all global illumination effects from shadows, and subsurface scattering, to reflections, refractions, and even caustics. For a successful emulation, we also need to obtain painterly versions of these global effects. As shown in this example, once we have a good estimate of scene description that can mimic the original information reasonably well, we can use this information even for stylistic emulation. 
\item \textbf{Hyper-Realistic Augmentations:} These can be defined as a process of rendering hyper-realistic images by adding the impact of the new objects in the scene. The original scene is represented using proxy cameras, proxy lights, proxy geometry, and proxy materials. The new objects are represented by virtual geometry and virtual materials. The impact of the virtual objects on the proxy scene (e.g. their shadows, their reflections, and their refractions) is computed as a new image. This new image is combined with the original background image to produce a single composite image \cite{abad2003,brinkmann2008,mitja2010}. The applications of augmentation are movie post-production and augmented reality. Figure~\ref{fig_GaryBruins} demonstrates the importance of global illumination in image augmentation. In this example, a background image (see Figure~\ref{fig_GaryBruinsa}) is combined with a foreground image of a rolling pin Figure~\ref{fig_GaryBruinsb}. The simple combination of two images that are rendered with the same camera parameters, as shown in Figure~\ref{fig_GaryBruinsb}, does not produce a believable perceptive image. Figure~\ref{fig_GaryBruinsc} shows a composite image with global illumination, which is more appealing. As shown in the visual comparison of these two examples, creating appropriate shadow, reflection, refraction, and caustic images and combining them into a background image is critical for achieving realistic compositing. 
\end{itemize}

\begin{figure}[htbp!]
    \centering
    \begin{subfigure}[t]{0.324\textwidth}
        \includegraphics[width=1.0\textwidth]{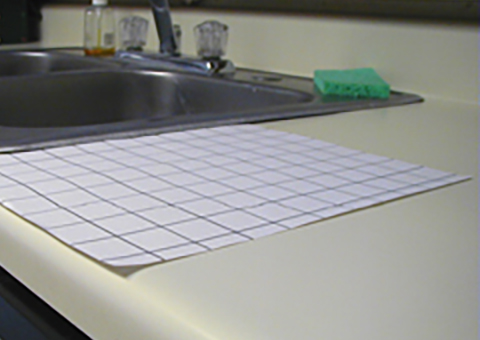}
         \caption{     Determining position and orientation of the tabletop.}
        \label{fig_GaryBruins2a}
    \end{subfigure}
    \hfill
    \begin{subfigure}[t]{0.324\textwidth}
        \includegraphics[width=1.0\textwidth]{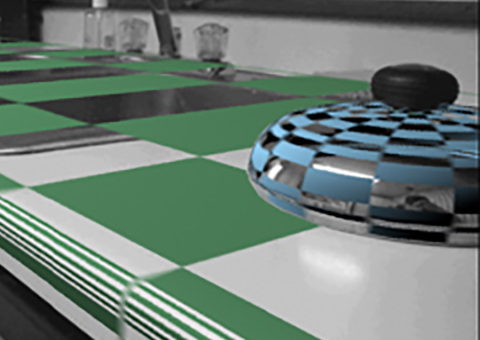}
         \caption{     Representation of Shapes and Positions of Objects.}
        \label{fig_GaryBruins2b}
    \end{subfigure}
    \hfill
    \begin{subfigure}[t]{0.324\textwidth}
        \includegraphics[width=1.0\textwidth]{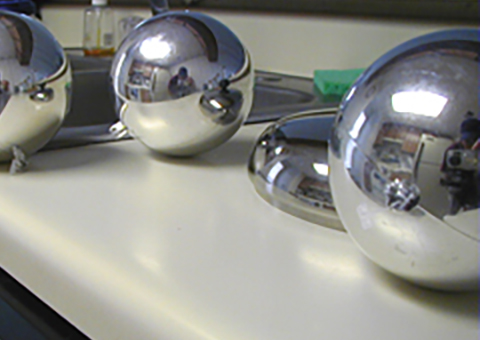}
         \caption{    Representation of Shapes and Positions of Lights.}
        \label{fig_GaryBruins2c}
    \end{subfigure}
     \caption{    To create shadow and reflection in the composited images shown in Figure~\ref{fig_GaryBruins}, there is a need to reconstruct the proxy shapes of the countertop \ref{fig_GaryBruins2a} and the lid \ref{fig_GaryBruins2b}; and the positions and light intensities \ref{fig_GaryBruins2c}.}
    \label{fig_GaryBruins2}
    \end{figure}

\subsection{Context \& Motivation}

Consider the augmentation case shown in Figure~\ref{fig_GaryBruins}. Note that the process consists of a series of at least three complicated operations: (1) production of the shadow of the computer-generated object (s), (2) production of reflection of the computer-generated rolling pin, and (3) production of an appropriate image of the rolling pin.  Each of these operations requires an appropriate representation of the positions, shapes, and material properties of all objects and lights in the background scene. This is a relatively simple case.  The real scenes usually have multiple transparent objects and there can be multiple distinctly colored lights.  Unfortunately, the inclusion of these effects for hyper-realistic illusions is prohibitively expensive in the current process for applications that require real-time interaction such as Augmented Reality(AR) or Virtual Reality(VR).

The main problem with current practice for real-time applications is that the process requires a reasonably good representation of real objects and implementation of fundamental physical laws of optics, both of which are prohibitively difficult and expensive. For example, the true 3D representation process demonstrated in Figure~\ref{fig_GaryBruins2b} is not trivial, since real objects are almost always imperfect. When we use proxy cameras, proxy lights, proxy geometry, and proxy materials, there will always be some relatively minor reconstruction errors that can cause significant disturbances in visual perception. Moreover, we need information for the whole, as demonstrated in Figure~\ref{fig_GaryBruins2c}, to use image-based lighting, which may not always be possible. Also, physically-based realistic rendering with global illumination can take a significant amount of time. In other words, although the creation of hyper-realist illusions is common \emph{especially in the movie industry}, it is still a laborious process that requires a significant amount of computing and human resources. This resource problem cannot be solved simply using brute-force solutions by using faster computers and better GPUs. \emph{There is a need for a significant paradigm shift to provide hyper-realism in real-time applications such as VR or AR.} 

\section{Previous Work}

Hyper-realist rendering can be considered to be related to
non-photorealistic rendering (NPR) in computer graphics. NPR emerged as a subfield of computer graphics during the 1990s to produce computer-generated images that invoke the appearance of being created "by hand" \cite{haeberli1990paint,saito1990comprehensible,strothotte2002non,gooch2001non} by emulating broad artistic styles such as outlines and silhouettes \cite{hertzmann1999silhouettes,northrup2000artistic,isenberg2002stylizing}, clip arts \cite{stroila2007clip}, technical illustrations \cite{gooch1998non,diepstraten2002transparency}, pen and ink drawings  \cite{winkenbach1994computer,salisbury1997orientable,markosian1997real,deussen2000computer,salisbury2023interactive}, impressionist  \cite{meier1996painterly,litwinowicz1997processing}, implicit painting \cite{akleman1998,akleman1998a}, and cubist paintings \cite{meadows2000a,smith2004,morrison2020remote,akleman2023recursive}, Chinese painting \cite{chan2002two,xu2003advanced,wang2007image,li2020real}, charcoals \cite{majumder2002real,du2016charcoal}, and stippling \cite{lu2002non,secord2002weighted,martin2017survey,galea2016stippling}; as well as artistic tools and mediums such as brush strokes \cite{vanderhaeghe2013stroke,yeh2002animals,hertzmann1998painterly,lin2012video}, watercolor \cite{curtis1997computer}. Convolution neural networks are also effective for style transfer \cite{gatys2016image,selim2016painting}. On the other hand, the inclusion of Global Illumination Effects in hyper-realist and painterly emulations has not been very common in computer graphics research. 

Barycentric shading has been useful for obtaining global effects in a wide variety of painterly styles \cite{akleman2016}. Liu created reflection effects for a dynamic and 3D version of a Jiangnan water country painting by the contemporary Chinese artist Yang Ming-Yi \cite{liu2015chinese}.  Zhao created subsurface effects on an ocean surface for an Ayvazovski painting \cite{yan2015}. Deng also obtained the reflection effect on animated water in Chinese paintings using post-processing operations \cite{deng2023digital}. Subramanian created painterly reflection, refraction, and caustics with a classical still-life painting of wine and glass \cite{subramanian2020painterly}. Du obtained global illumination effects with charcoal and cross-hatching \cite{du2016charcoal,du2017designing}.   Clifford created painterly reflections with moving water in a dynamic painting emulating a Venice painting of Anne Garney \cite{clifford2021}. Ross controlled the color in the shadow regions with Bezier surfaces in a shader emulating the style of Georgia O’Keeffe \cite{ross2021Georgia}. Castenada obtained the "painterly" subsurface scattering effect for human faces \cite{castaneda2017paper}. Global illumination effects can also be obtained with limited information only with vector fields and images \cite{wang2014,wang2014global,akleman2017}. 
Using these methods, it is also possible to obtain interactive global illumination effects in real-time with the desired artistic styles \cite{akleman2023webbased}.

\section{Theoretical Framework}

Our approach to obtaining hyper-realism is based on the physical limitations of artists in collecting a complete knowledge of a scene. Artists do not have complete 3D shape information for all objects in the scene. They have only partial knowledge of geometry from a limited set of points of view. Artists also do not know the complete reflectance distribution functions. Again, they have only partial knowledge of reflectance distribution functions from a limited set of points of view. 

Based on a straightforward interpretation of these limitations, painting or illustration can appear to be post-production processes that are applied to a photograph or a rendering.\footnote{In fact, many non-photorealistic approaches, such as \cite{haeberli1990paint,saito1990comprehensible}, are essentially post-processing applications.} However, painters or illustrators do not view the scenes from only one point of view. Although they do not access full 3D shapes of all the objects in the scene, they can still move their head to get a better idea about the shapes. Moving their heads, they can also identify different global effects, they can understand the shapes of the silhouettes, they can understand the depth with parallax, etc.  

In other words, although painting and illustration are essentially post-production processes, they are based on not only "photographs" but also a deeper knowledge of the scenes that can be dynamically observed. In some sense, the visible world can conceptually be decomposed into regions by artists. Each region is processed differently based on its material properties. On the basis of this observation, we present a new conceptual framework to produce hyper-realist results. This conceptual framework allows us to decompose the hyper-realist rendering process into "illumination" and "shading". In other words, we view the rendering process in two stages called "Illumination" and "Shading". We consider illumination as a process of building realistic and non-realistic images, such as photographic images or paintings. These images can be relatively simple, missing rich details. On the other hand, we view shading as a process to convert these simple or uninteresting images into hyper-realistic, complex, and interesting images.

\begin{figure}[htbp!]
    \centering
    \begin{subfigure}[t]{0.59\textwidth}
        \includegraphics[width=1.0\textwidth]{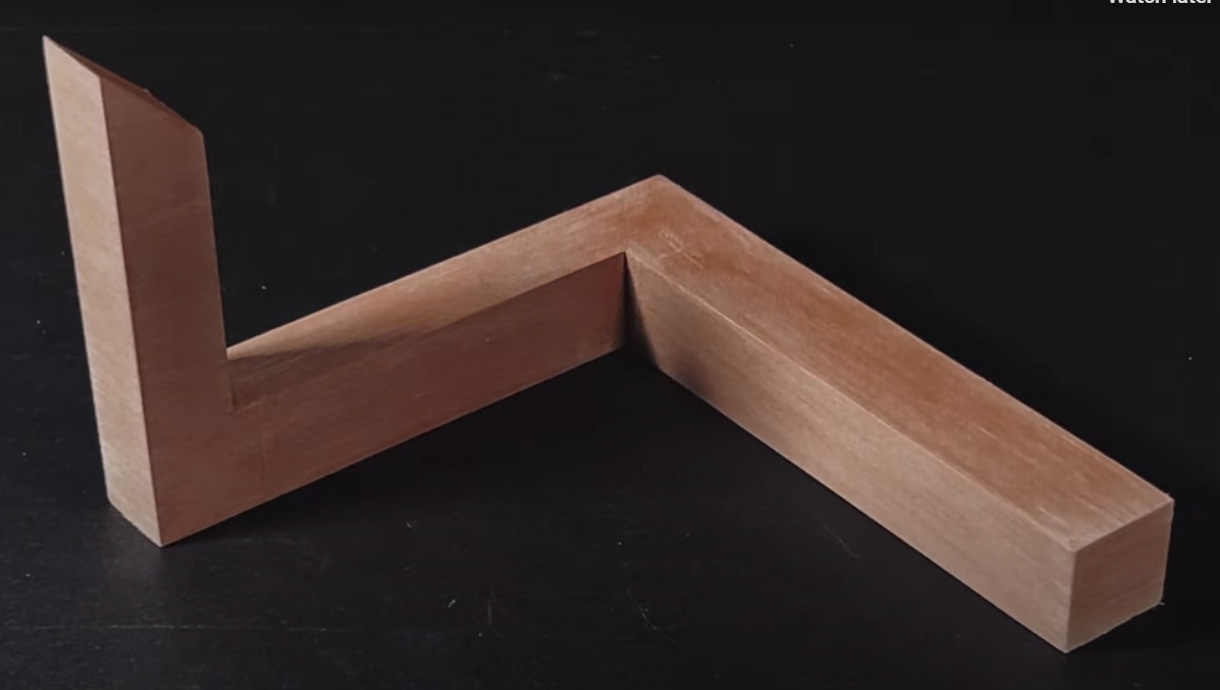}
         \caption{A shape sculpted by wood.}
        \label{fig_anamorphic/i1}
    \end{subfigure}
    \hfill
    \begin{subfigure}[t]{0.40\textwidth}
        \includegraphics[width=1.0\textwidth]{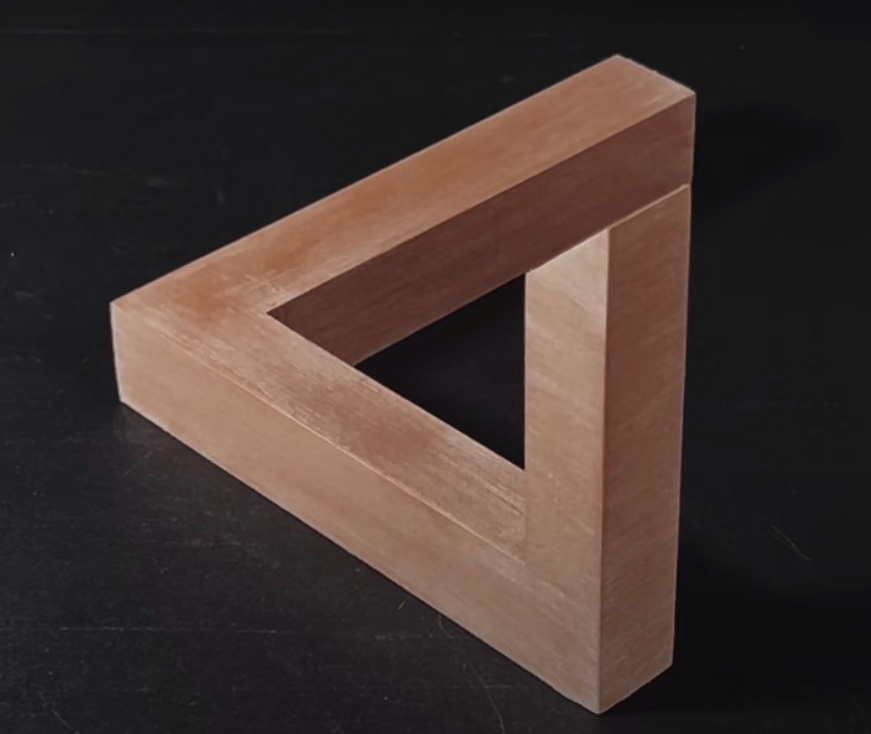}
         \caption{Perspective view from right vantage point creates an illusion of impossible object.}
        \label{fig_anamorphic/i0}
    \end{subfigure}
     \caption{This figure shows how anamorphism can be used to create illusion of impossible objects.  It is important to note that there exist infinitely many solutions that can produce the same impossible object \cite{Elber2011}.  }
    \label{fig_anamorphic1}
    \end{figure}

A key property of our framework is to use non-realistic geometry with basic materials to compute illumination and shading to support impossible, inconsistent, and incoherent shapes \cite{akleman2024representing, akleman2024compositing} (see Figure~\ref{fig_anamorphic1}). In other words, these geometric shapes and basic materials are allowed to form non-Riemannian manifolds \cite{bujack2022non, akleman2023circular}. This is a very critical property to obtain artistic results, since artistic creations often include impossible, inconsistent, and incoherent elements. An additional advantage of our framework is that we can use simpler geometry and materials, since geometry and materials do not have to satisfy the Riemannian property \cite{gromov1999metric,bujack2022non}. 

We need to point out a caveat in this way of thinking. Almost all shapes used in computer graphics are not Riemannian manifolds, since they are not smooth, that is, they are not infinitely differentiable \cite{lee2018introduction}. For example, triangular meshes are not really differentiable in their edges and vertices. We need to clarify that by the Riemannian property, we mean that the property of the distance between two points on the manifold is the length of the shortest path that connects them. This can be uniquely defined by adding small differences in any given geodesic \cite{gromov1999metric,bujack2022non}.. This property suggests that if we travel along a path, we can return to the original position, and only some of the small differences become zero. This property can hold even when the shapes are not Riemannian manifolds. However, the property must not hold in impossible, inconsistent, or incoherent shapes. For example, consider the never-ending staircase in M. C. Escher's famous \textit{Ascending and Descending} lithograph \cite{schattschneider2010mathematical}. In that lithograph, two groups of identically dressed men are continuously ascending and descending the never-ending staircase \cite{escher1960ascending}. Since we cannot return the same height by continuously climbing, this is an impossible shape and obviously does not satisfy the Riemannian property. Note that a good example of such structures is non-conservative vector fields \cite{akleman2024representing}.  Gradient fields, on the other hand, can satisfy the Riemannian property \cite{akleman2024representing}. 

\begin{figure*}[htb]
  \centering
        \includegraphics[width=1.00\textwidth]{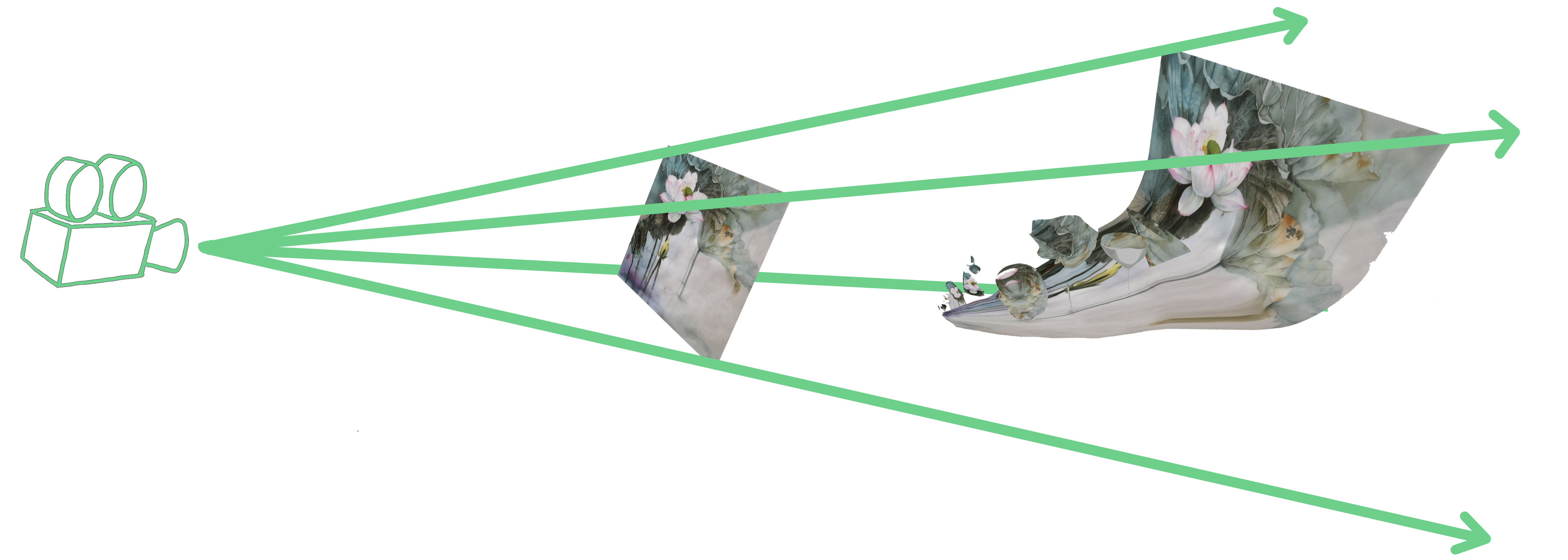} 
\caption{Anamorphism is obtained by simply projecting a 3D-looking image of an object or scene from a vantage point to a random 3D geometry. Viewing the projected image from the same vantage point creates the illusion of a 3D object or scene.  }
\label{fig_anamorphism2}
\end{figure*}

Non-conservative vector fields are useful for representing shapes that do not satisfy the Riemannian property \cite{akleman2024representing, akleman2024compositing}. Fortunately, vector fields are not the only ones. There is another representation that can support the non-Riemannian property. Consider the impossible object in Figure~\ref{fig_anamorphic/i0}. The shape in Figure~\ref{fig_anamorphic/i1} is just a normal shape that can be considered a Riemannian 2-manifold if we assume that it is infinitely differentiable at the sharp corners. On the other hand, it appears to be an impossible object \ref{fig_anamorphic/i0} from a vantage point. Note that this shape from the vantage point can be considered as a height field. This height field is discontinuous in one region and, therefore, is not-differentiable. In other words, we can also produce impossible shapes with height fields by introducing discontinuities from a vantage point. This corresponds to an anamorphic representation of height fields (see Figure~\ref{fig_anamorphism2}). As a result, this gives us two types of representations that can be used for hyper-realistic illumination. 

Based on this viewpoint, we can start by computing illumination using such relatively simple non-Riemannian geometries to obtain relatively simple images. Then, by converting these relatively simple images with stylistic reflectance distribution functions, we obtain a final hyper-realist rendering. The process is conceptually simple if there is no constraint, i.e. if we create our artistic images. On the other hand, if we use this process for emulations or augmentations, the problem is an inverse problem. We need to be careful, especially with registration problems.  

For illumination computation, we need (1) simplified non-Riemannian geometric representations of all objects that appear to be correct only from a limited set of points of view; and (2)  reflectance distribution functions that are roughly acceptable for most invisible points. On the other hand, for shading computation, we need some type of stylistic reflectance distribution maps (functions) for all visible points that can convert the original illumination images into new and more stylistic images. Fortunately, there are already some methodologies to represent such simplified models.  In the rest of the paper, we give details of these illumination and shading stages.

\begin{figure}[htbp!]
    \centering
    \begin{subfigure}[t]{0.485\textwidth}
        \includegraphics[width=1.0\textwidth]{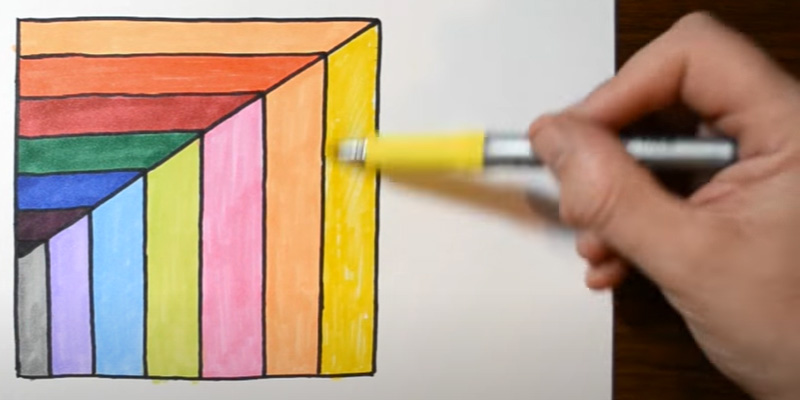}
         \caption{The top view of a simple image.}
        \label{fig_anamorphic/0}
    \end{subfigure}
    \hfill
    \begin{subfigure}[t]{0.485\textwidth}
        \includegraphics[width=1.0\textwidth]{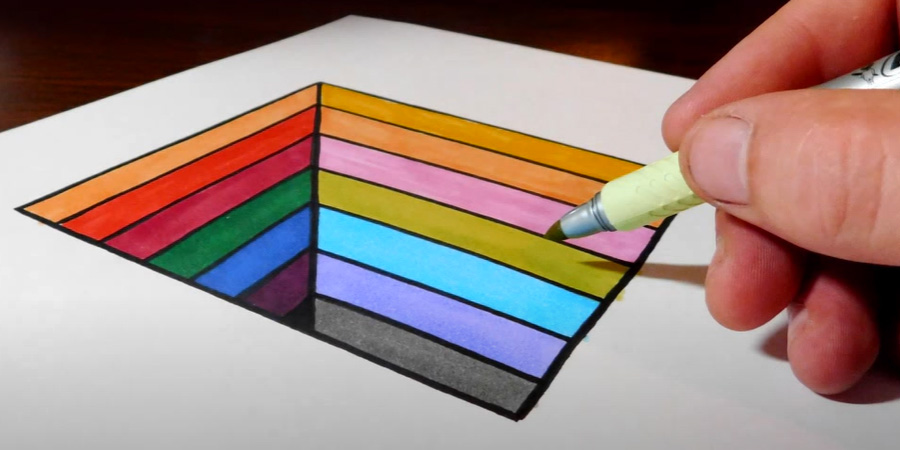}
         \caption{Perspective view from right vantage point creates an illusion of a square hole.}
        \label{fig_anamorphic/1}
    \end{subfigure}
    \hfill
        \begin{subfigure}[t]{0.485\textwidth}
        \includegraphics[width=1.0\textwidth]{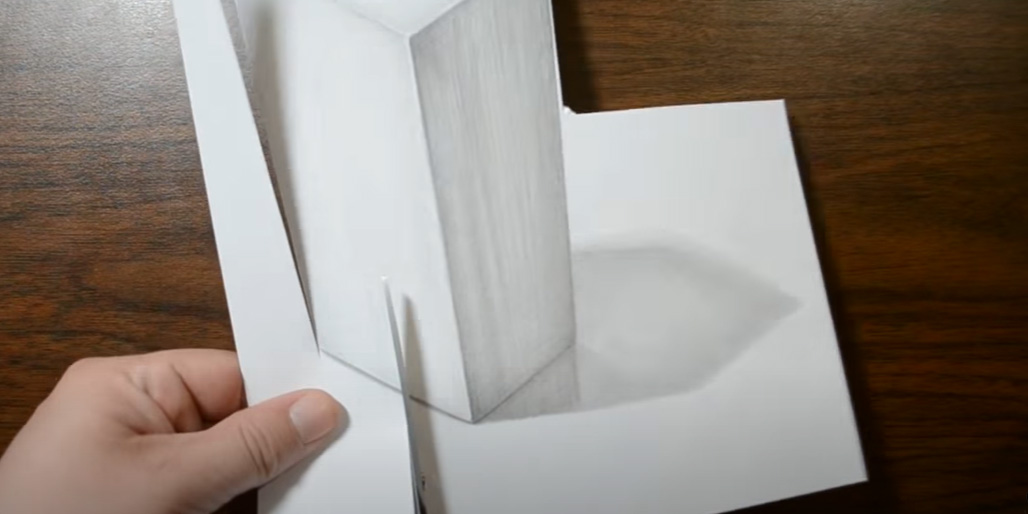}
         \caption{The top view of an exaggerated drawing.}
        \label{fig_anamorphic/2}
    \end{subfigure}
    \hfill
        \begin{subfigure}[t]{0.485\textwidth}
        \includegraphics[width=1.0\textwidth]{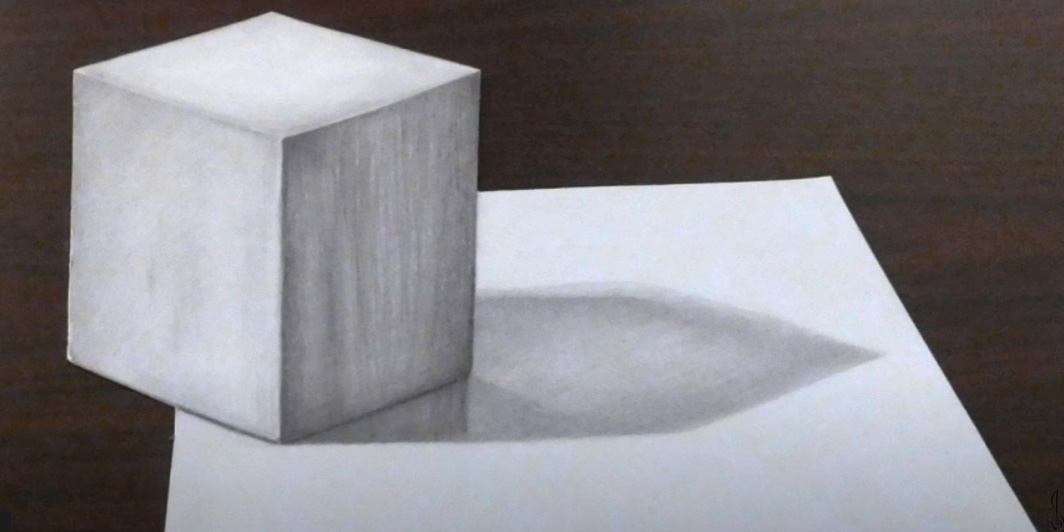}
         \caption{Perspective view from right vantage point creates an illusion of a 3D cube .}
        \label{fig_anamorphic/3}
    \end{subfigure}
    \hfill
     \caption{Examples of Anamorphic images that are obtained as screen-captures of video instructions by Jon Harris \cite{harris2017drawing,harris2018how}, who demonstrates how easy it is to create anamorphic art even for children. }
    \label{fig_anamorphic}
    \end{figure}

\subsection{Illumination Stage: Rendering with Simplified non-Riemannian Geometry \& Basic Materials as Virtual and Proxy Objects}

As described in the previous section, in the hyper-realist illumination process, we represent both virtual and proxy objects with relatively simple non-Riemannian geometries and basic materials to support impossible, inconsistent, and incoherent structures. In this section, we briefly describe how to obtain and use such structures. First, we point out that, in computer graphics, all geometries and materials are essentially simplified representations that do not necessarily support the Riemannian property. However, the omission of the Riemannian property is usually practical and is not necessarily intentional. For example, we do not check normal maps if they are gradient fields or not. Most likely, many of the normal maps do not satisfy the Riemannian property, but since they somewhat work, no one enforces theoretical consistency. In hyper-realism, we theoretically embrace the non-Riemannian property. In this section, we briefly discuss how to build simple non-Riemannian geometries and materials that can support hyper-realist emulations and augmentations. 

The simplest geometries that can be used as proxy and virtual object shapes are the Mock-3D shapes, which appear to be 3D but are essentially 2D \cite{wang2014qualitative}. In mock-3D shapes are essentially a combination of two art approaches: anamorphic art \cite{schulman2020anamorphic, odeith2024anamorphic, beever2024apavement} and bas-reliefs \cite{belhumeur1999bas, weyrich2007digital}. The advantage of both types of representation is that they can directly support impossible, incoherent, and inconsistent structures. In anamorphic art, the perspective information is embedded into the picture by projecting the picture into a 3D scene. Therefore, the anamorphic images appear to be 3D from a vantage point. Figure~\ref{fig_anamorphism2} shows the main approach to obtain anamorphic art.

Bas-reliefs, on the other hand, are scaled-down geometries \cite{belhumeur1999bas, weyrich2007digital} that are qualitatively similar to actual geometry (see \cite{wang2014qualitative} for a discussion). Anamorphisms are often included with bas-reliefs even in early cave art to create the illusion of 3D \cite{akleman2024representing}. We call these Mock-3D shapes anamorphic bas-reliefs. Note that anamorphism and bas-reliefs are exactly how we can create 3D impossible shapes in practice \cite{Elber2011}. 

\begin{figure}[htbp!]
    \centering
    \begin{subfigure}[t]{0.45\textwidth}
        \includegraphics[width=1.0\textwidth]{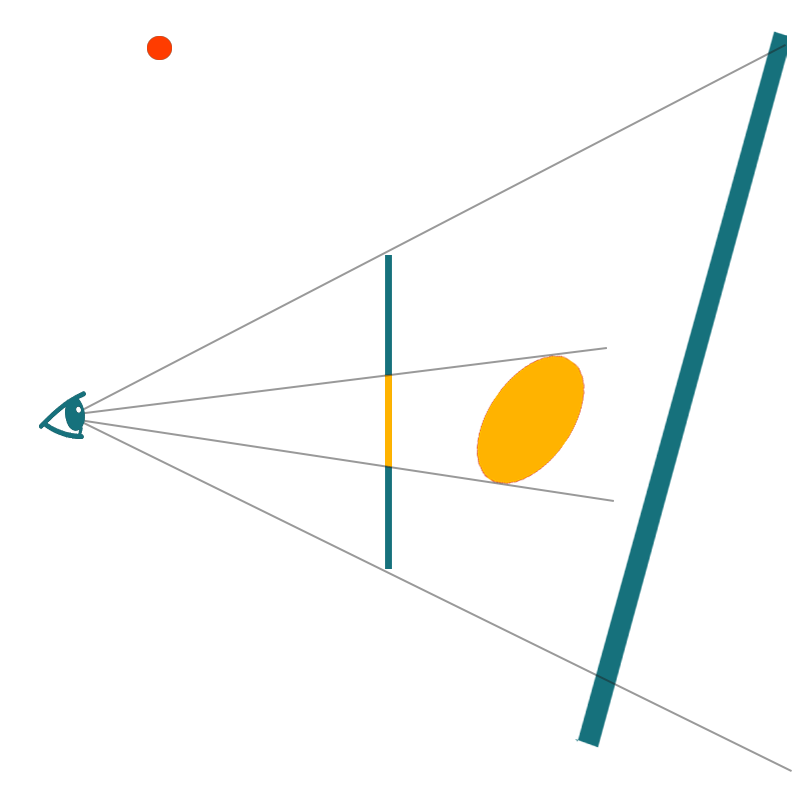}
        \caption{ A flatland scene. The red circle is a light. The cyan rectangle and yellow ellipsoid represent objects in the scene. The eye is the camera. The line is the 1D image.  }
        \label{fig_embedded0}
    \end{subfigure}
    \hfill
    \begin{subfigure}[t]{0.45\textwidth}
        \includegraphics[width=1.0\textwidth]{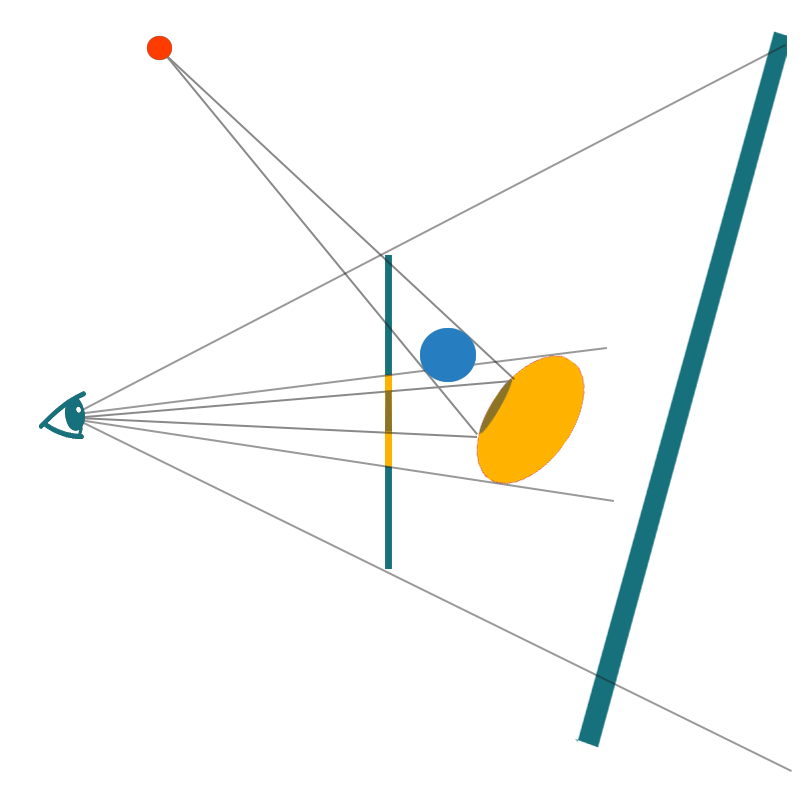}
        \caption{ Expected shadow of a 2D CG object in Flatland scene. The CG object is the blue circle. }
        \label{fig_embedded2}
    \end{subfigure}
    \hfill
    \begin{subfigure}[t]{0.45\textwidth}
        \includegraphics[width=1.0\textwidth]{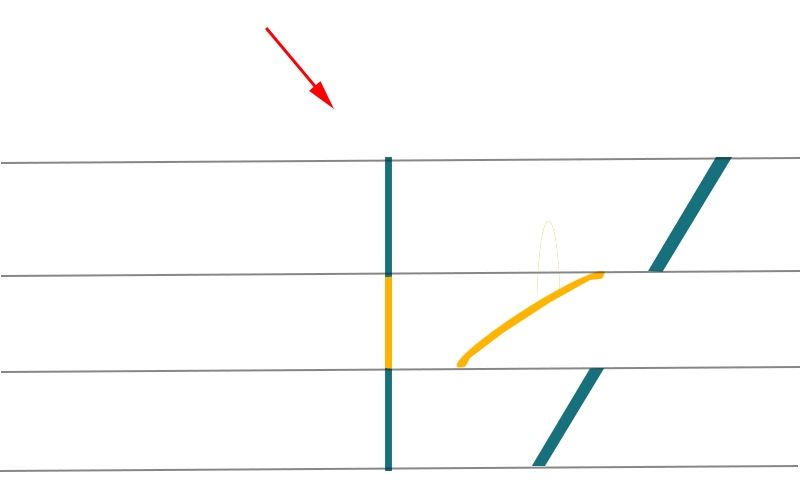}
        \caption{ 1D Proxy shapes that are used to reconstruct the flatland scene. The original light is replaced by a directional light. The original objects are replaced by qualitative similar curves that are given as a height function. The perspective is embedded in shapes by using a parallel projection. Note that the image is still the same.}
        \label{fig_embedded1}
    \end{subfigure}
        \hfill
    \begin{subfigure}[t]{0.45\textwidth}
        \includegraphics[width=1.0\textwidth]{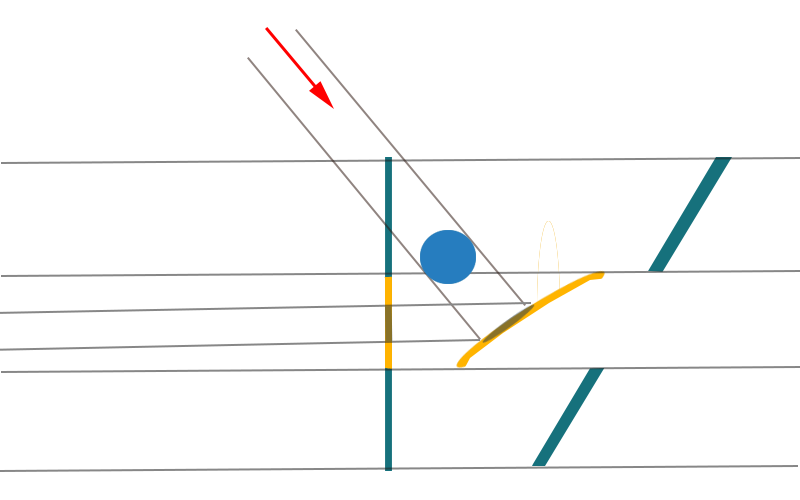}
        \caption{The shadow of a CG object on proxy objects. Note that the shadow is similar to the original shadow mainly because the proxy objects are qualitatively similar to the original objects.  }
        \label{fig_embedded3}
    \end{subfigure}
    \hfill
    \caption{A 2D Flatland example that demonstrates an anamorphic bas-relief that represents a complicated scene with qualitatively similar proxy shapes that embed perspective transformation can provide qualitatively acceptable global illumination effect. This example visually shows that it is possible to obtain shadows that are reasonably close to original shadows. We do not even reconstruct light positions accurately. }
    \label{fig:020embedded}
\end{figure}

Note that bas-reliefs do not have to be represented by 3D geometry. We can even use normal maps \cite{Johnston2002, Okabe2006, Shao2012}. One problem with normal maps is that they are supposed to be gradient fields and therefore they cannot theoretically represent impossible, incoherent, and inconsistent shapes such as those used by artists such as Pablo Picasso of M. C. Escher \cite{akleman2024representing}. This is not really a major problem. Normal maps can support non-conservative vector fields and can be used directly to represent impossible, incoherent, and inconsistent shapes \cite{wang2014global,wang2014qualitative}. The main problem is that normal maps do not have thickness information, which can create problems with thin objects or representing transparent objects \cite{akleman2024compositing}. The thickness information can also be embedded in normal map images, in an extended form of normal maps, called shape maps \cite{akleman2024representing,akleman2024compositing}. 

\begin{figure}[htb!]
  \centering
        \includegraphics[width=1.0\textwidth]{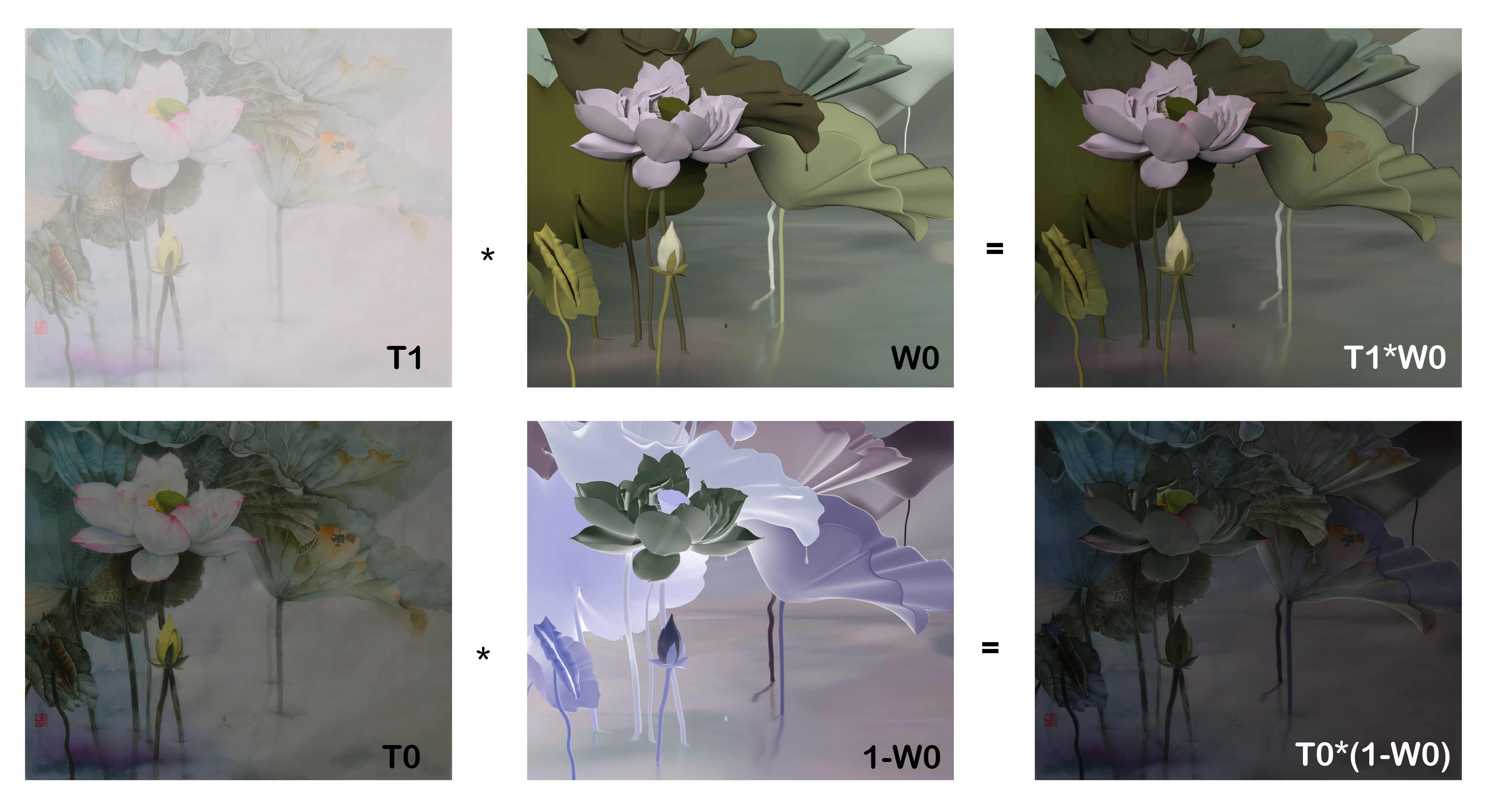}
        \includegraphics[width=1.0\textwidth]{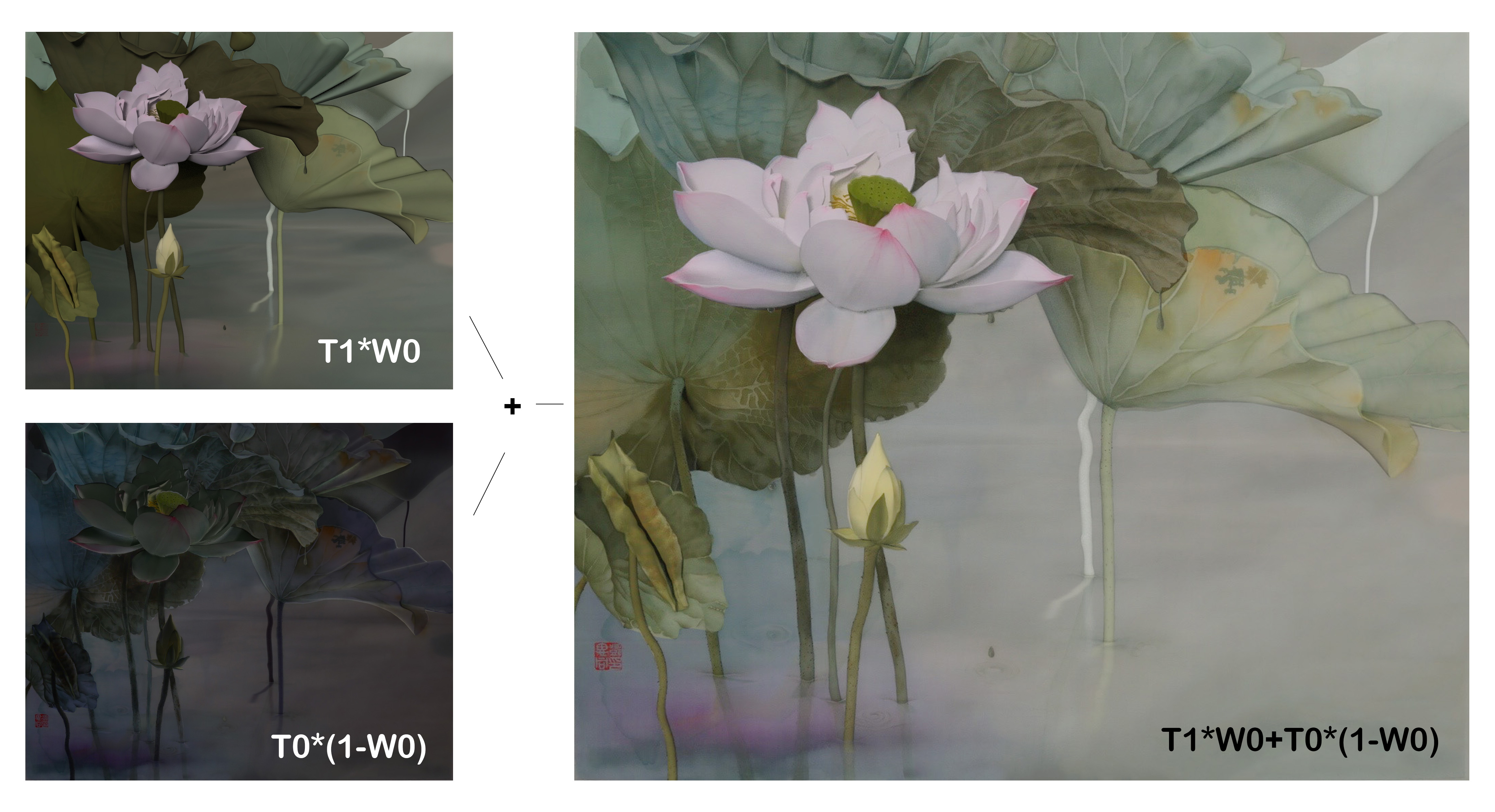}
\caption{Compositing operation with a manipulated $W(u,v,t)$ image.}
\label{fig_compositing0}
\end{figure}

Anamorphic bas-reliefs are mock3D shapes with associated materials. An additional advantage of anamorphism is that it makes the assignment of materials with texture mapping simple, just a perspective projection. Such objects can essentially be formulated as a layer that is represented as a parametric surface that is given as $\mathbf{P}(u,v)$ where $(u,v)$ are parametric coordinates in $[0,1]^2$. The simplest of such layers is a planar rectangle, and in most cases, it is sufficient to use a planar rectangle. We can also use tensor product B-spline surfaces or simply a wrinkled polygonal mesh. Since these are always functions of a square domain $[0,1]^2$ to the 3D space $\mathbf{P}=(x,y,z)$, it is easy to texture map them. We can simply use the parametric coordinates $(u,v)$ as texture coordinates. Without loss of generality in the rest of the paper, we assume that it is a unit square in $z=0$ as $\mathbf{P}(u,v)=(u,v,c)$ where $(u,v) \in [0,1]^2$ and $c$ is a constant, which is usually zero. 

\begin{figure}[htb!]
  \centering       \includegraphics[width=1.0\textwidth]{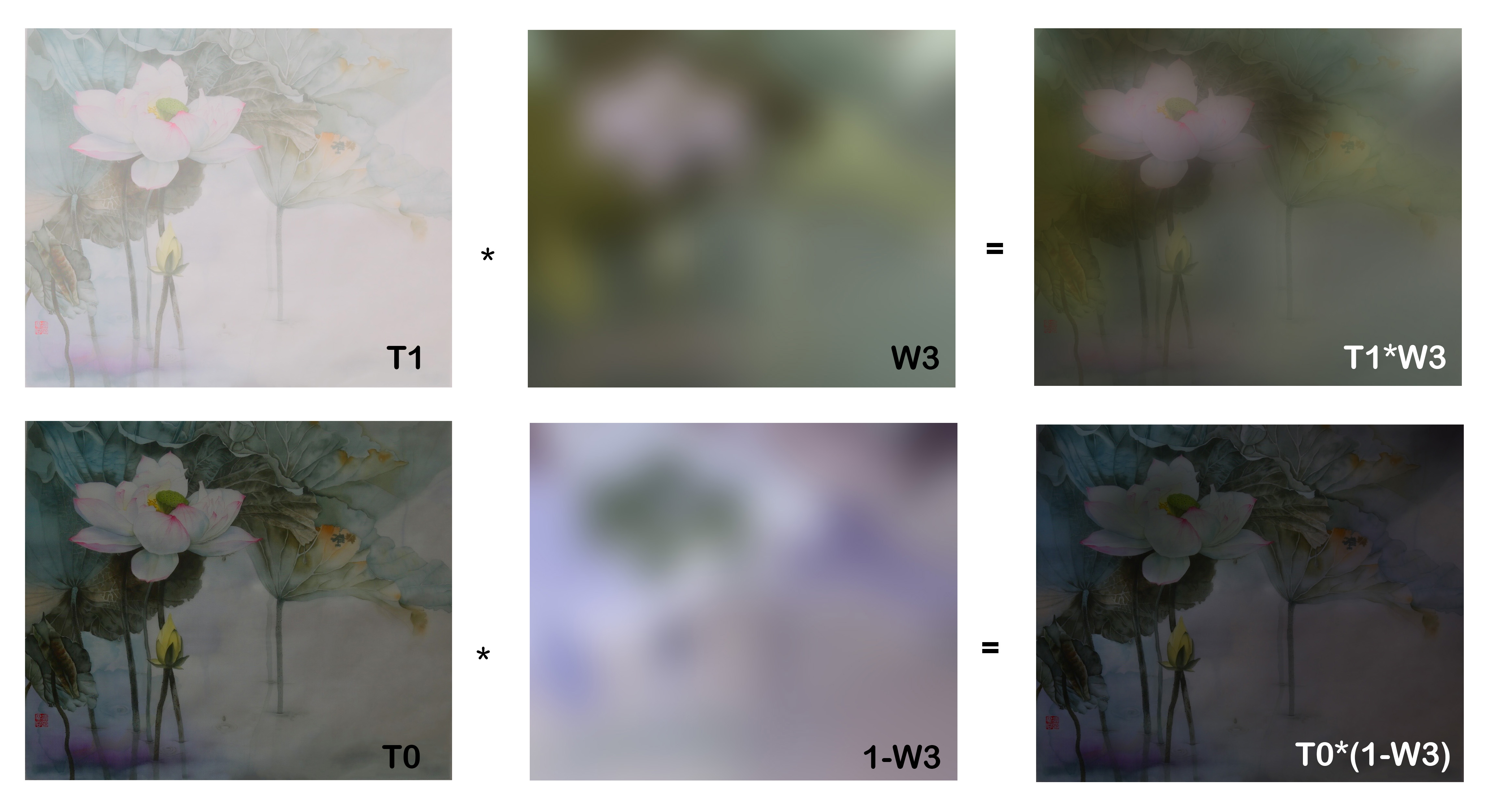}       \includegraphics[width=1.0\textwidth]{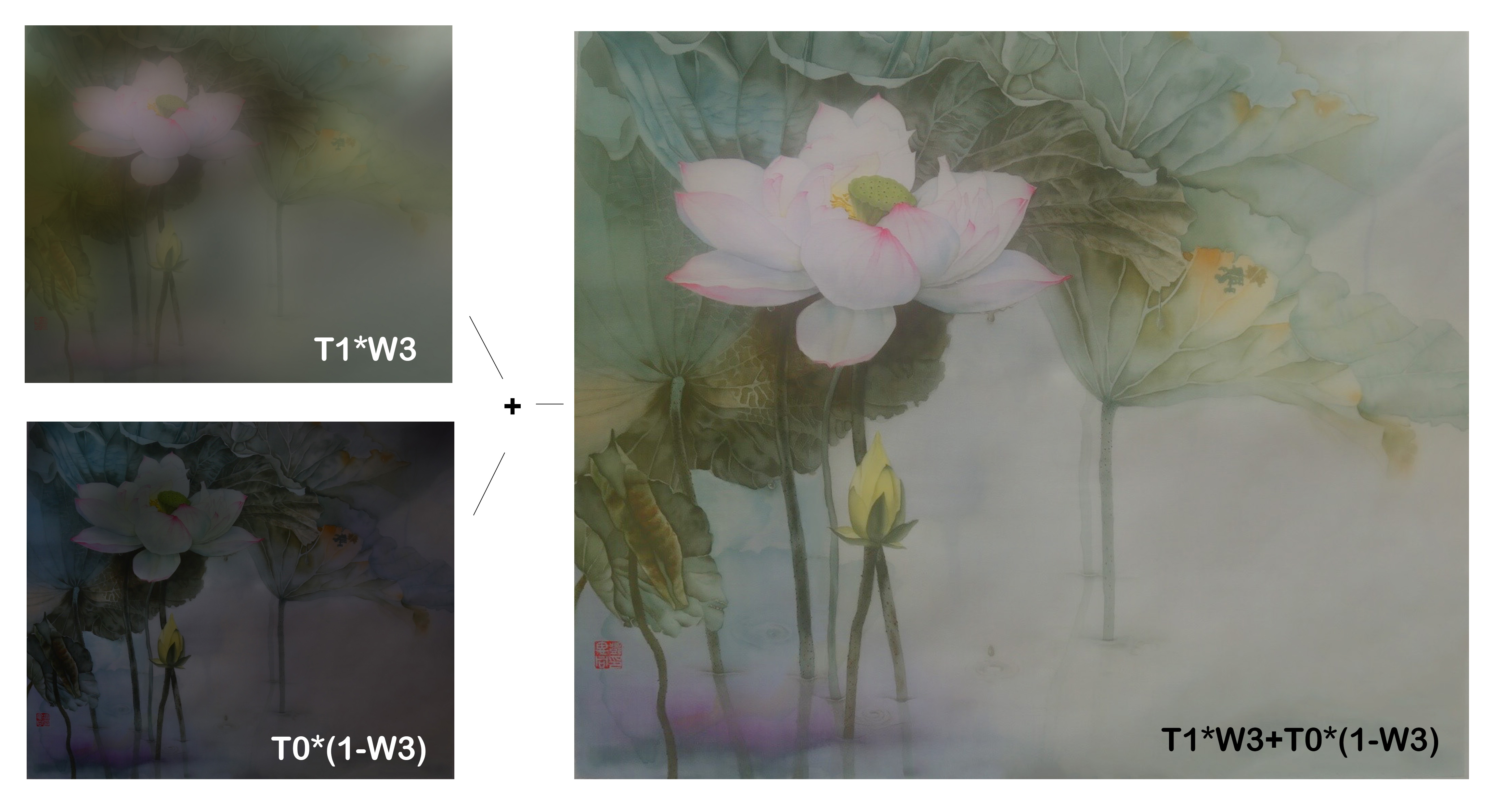}
\caption{Compositing operation with a manipulated $W(u,v,t)$ image. In this case, we blurred $W(u,v,t)$ to obtain show that the result is robust. Note that the term that corresponds to the classical (non-barycentric) diffuse formula $T_1W_3$ is just a blurred image. On the other hand, the barycentric formula still provides clean results. }
\label{fig_compositing3}
\end{figure}

Each of these rectangular texture layers can consist of several components, each of which is a texture projected onto the rectangular surface. We call these components ``channels'' consistent with the standard terminology used in image manipulation. However, in this case, the term channel will refer to entities that are more general than simple color channels. One channel provides proxy shape information, which we call Mock3D shapes, which are either height fields or vector fields. Height fields are represented by images called depth maps, and vector fields are represented by images called normal maps. The additional channels are control images that provide material properties of the objects.

An important property of using anamorphic bas-reliefs as virtual and proxy shapes is that embedding perspective transformations does not hurt computation of global illumination effects. We want to demonstrate how the concept works in a lower dimension shown in Figure~\ref{fig_embedded0}. In this particular case, we provide a scene from Flatland, that is, a 2D world \cite{abbott2006}. A flatlander's view of the world is 1D, i.e. one dimension lower, as shown in this figure. Using the same analogy, note that our view of the world is just 2D, again one dimension lower. If we keep the boundaries the same, we can obtain the same 1D image even using a parallel projection as shown in Figure~\ref{fig_embedded1}.   

To support more general scenes, mock-3D scenes can be constructed by projecting each set of images onto 2-manifold or 2-complex meshes \cite{akleman1999guaranteeing, akleman2000new, akleman2003minimal, akleman2024representing}. We can freely deform these shapes in the z direction to represent different versions of the same impossible object. This is particularly useful since we can deform objects without changing the visible parts \cite{Elber2011}.  Deformations at the depth of Z can also provide local layering \cite{mccann2009local}. Using 2-manifold meshes is standard computer graphics practice. for 2-complexes, see \cite{akleman2024representing}. 

\begin{figure}[htbp!]
    \centering
    \begin{subfigure}[t]{0.485\textwidth}
        \includegraphics[width=1.0\textwidth]{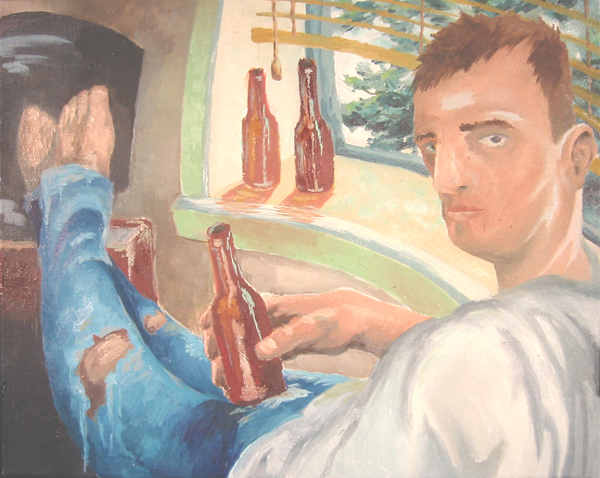}
         \caption{A figurative painting by Jonathan Kiker.}
        \label{fig_augment/afternoon1}
    \end{subfigure}
    \hfill
    \begin{subfigure}[t]{0.485\textwidth}
        \includegraphics[width=1.0\textwidth]{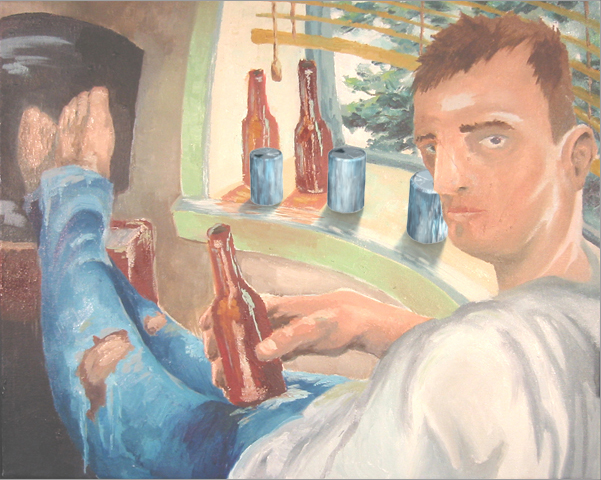}
         \caption{An augmentation of the painting with glasses.}
        \label{fig_augment/afternoonF}
    \end{subfigure}
    \hfill
     \caption{An example of hyper-realist augmentation of a figurative painting \cite{kiker2010using}. }
    \label{fig_augment/afternoon}
    \end{figure}

An important property of these 2-manifold meshes and 2-complexes is that they are not unique. As long as the edges of the silhouette match, the shapes are acceptable, such as the one shown in Figure~\ref{fig_anamorphic/i1}. This property makes the modeling process very simple. Another important property is that the textures that describe materials can be very simple, even a simple color. In the process of illumination, we only need some color bleeding. This even affects the color of the shadows, which is never constant. It is also important to estimate the approximate luminance and hue of the occluded regions \cite{akleman2023circular}. 

We point out that this type of representation is useful only from a vantage point of view. In hyper-realist rendering, we are philosophically against spending time to obtain very precise 3D scenes. If we have a high-quality 3D representation, they can work perfectly in this framework. On the other hand, many of the infinitely many potential anamorphic bas-reliefs can work to obtain basic illumination. We call this basic illumination, but, in the illumination stage, we can compute all global illumination effects. We simply ignore perceptually unimportant secondary effects. For instance, consider a wall made of slightly different colors of bricks, which works as a large reflector and illuminates the scene. In this stage, we do not need to consider all individual bricks. Their average color simply contributes to the color bleeding in the scene. For global illumination, we mainly need effects such as color bleeding in the scene. On the other hand, we need local shape information to compute local shadows \cite{wang2014global,wang2014qualitative,akleman2017,akleman2023web}. 

An important detail of the illumination stage is that the final rendered image, denoted by $W$ (see $W_0$ and $W_1$ Figures~\ref{fig_compositing0} and \ref{fig_compositing3}), is created as the total illumination of all visible points. Ignoring material properties of visible points is necessary since $W$ images will later be used to construct Hyper-realist images through the shading stage. If we include shader effects in the illumination stage, the shaders will be applied twice, which is incorrect. Ignoring material properties of visible points is straightforward for most of the ``backward'' rendering algorithms, such as path tracing, since they start with visible points. The only change in the algorithm is to ignore the first shader effect. It is also better to separate diffuse and specular illumination since shaders can act differently for these two \footnote{note that $W_0$ and $W_1$ Figures~\ref{fig_compositing0} and \ref{fig_compositing3} include both specular and diffuse parts.}.

An additional computation in the illumination stage is done to compute dynamic shaders that will be used in the shading stage. A critical idea is that, in the illumination stage, we also need to identify shader information at any given point. For example, consider the reflection of a brick wall we mentioned earlier. In the shading stage, we need texture information to see individual bricks on a mirror surface. Therefore, not only the illumination image $W$, but also the high-quality shading parameters $T_0$  and $T_1$ in Figures~\ref{fig_compositing0} and~\ref{fig_compositing3} are computed as texture images during the illumination stage. 

\subsection{Shading Stage: Stylistic Materials in Screen Space}

The concept of creating complex styles with images, which can be considered a high level of shader, is introduced in \cite{akleman2024compositing} and \cite{akleman2023recursive}. In this paper, we position those ideas into a bigger framework under the umbrella of hyper-realism. Figures~\ref{fig_compositing0} and~\ref{fig_compositing3} show the simplest use of this shading approach. In these cases, we use the shading stage with only diffuse materials. In this particular case. we use barycentric diffuse shaders \cite{akleman2016} to compute the final images. The shading stage is critical to obtaining desired styles for hyper-realist rendering. 

Figures~\ref{fig_compositing0} and~\ref{fig_compositing3} demonstrate that the quality of illumination is not that critical to obtaining the desired style. Insofar as illumination provides an appropriate average color in any given region, it is sufficient to use it. 
These are simple weighted averages of the texture images $T_0$ and $T_1$ using the illumination image $W$. The main advantage of barycentric shading is its robustness, i.e. the resulting rendering image is not ruined by the problems in illumination computation. In the particular case in Figure~\ref{fig_compositing0} reflection on the surface of the water is calculated by assigning the appropriate texture information to the surface of the water (see \cite{deng2023digital} for details). Many simple cases that only include mirror-like water can be implemented in this way, such as a 3D version of a Jiangnan water country painting by the contemporary Chinese artist Yang Ming-Yi as a primary visual reference \cite{liu2015chinese}.  The painting mainly uses diffuse surfaces such as Edgar Payne's landscapes can also be obtained by similar - i.e. relatively simple- diffuse barycentric shading \cite{justice2018}.

\begin{figure}[htbp!]
    \centering
    \begin{subfigure}[t]{0.485\textwidth}
        \includegraphics[width=1.0\textwidth]{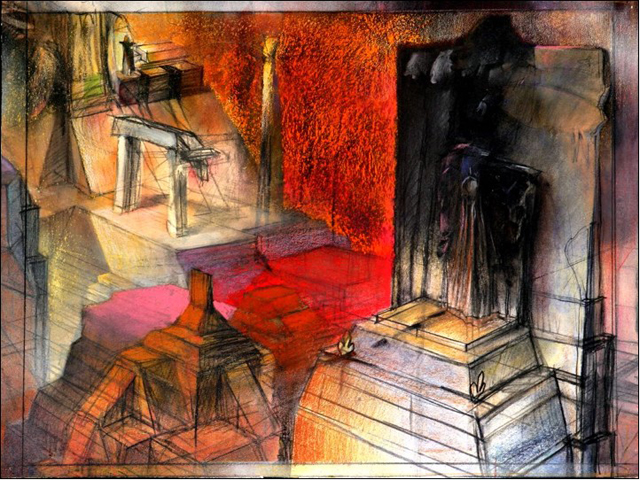}
         \caption{A figurative painting by Dick Davison.}
        \label{fig_augment/d13}
    \end{subfigure}
    \hfill
    \begin{subfigure}[t]{0.485\textwidth}
        \includegraphics[width=1.0\textwidth]{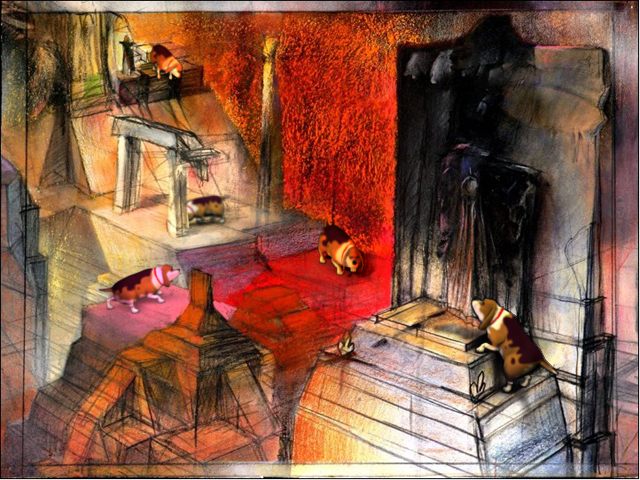}
         \caption{An augmentation of the painting with dogs.}
        \label{fig_augment/d13F}
    \end{subfigure}
    \hfill
     \caption{Another example of hyper-realist augmentation of a figurative painting \cite{kiker2010using}. }
    \label{fig_augment//davison13}
    \end{figure}

On the other hand, if we want to have more complex effects such as reflection and refraction, nonlinear diffuse color change, subsurface scattering, or caustics, bush stroke effects need more complicated barycentric shading formulations. For example, Ross used high-dimensional Bezier functions to emulate the style of Georgia O'Keeffe that requires non-linear color changes in diffuse illumination \cite{ross2021Georgia}. Another set of difficult examples of landscape paintings. For example, the sea paintings of Ivan Aivazovsky include a high-quality subsurface scattering with water that requires a more complicated barycentric shader \cite{yan2015}. The classical still life paintings require a barycentric shader that can support painterly reflection, refraction, Fresnel effects and caustics \cite{subramanian2020painterly}. Moreover, if we want to obtain charcoal, hatching, or cross-hatching, we need to use a barycentric shader with a zero degree B-spline \cite{du2017designing,du2016charcoal}. 

Regardless of complexity, shaders should support an algebraic structure to obtain all the colors that are necessary to obtain the desired style.  Barycentric shaders are desired since they provide such an algebraic structure \cite{akleman2016}. However, we observe that barycentric methods are not the only ones that obtain algebraic structures. For example, it is also possible to represent colors with complex numbers to support an algebraically consistent structure \cite{akleman2023circular}.

\section{Conclusion and Discussion}

Despite significant technical advances in real-time applications, obtaining high-quality visuals is still not practical in most real-time scenarios. Most of the current interactive real-time experiences present visuals that lack the basic elements necessary to obtain a realistic perception of a scene (similar to the example shown in Figure~\ref{fig_GaryBruinsb}). Our goal is to eventually achieve hyper-realism in real-time by presenting visuals that are of a quality comparable to the image shown in Figure~\ref{fig_GaryBruinsc}, including all global illumination effects from shadows to reflection, refraction, and caustics with art-directable visuals. These effects are critical to improving user cognition and performance while providing non-ambiguous visual perceptions of reality, as even a small detail such as the position of a shadow can alter our perception of the environment.

We claim that representational art-based approaches provide the type of simplified methods to obtain non-ambiguous visual perception in real-time applications. In current practice, the predominant use of hyper-realism is in the motion pictures industry and is achieved through a laborious and time-consuming post-production process. These hyper-realist methods can still keep humans in the loop and provide a rich toolbox for developers, designers, and artists to fulfill distinct visual goals. This toolbox will enable users to (1) re-purpose the background scene to improve an AR experience for individual needs, and (2) achieve effects that are essential for visual perception, including critical global illumination components such as shadows, reflection, refraction, and caustics. 

Our primary achievement in this paper is to provide a theoretical framework to simplify shapes and materials to support hyper-realism in real-time interactive applications such as VR or AR. Our secondary goal is to understand how to tailor these shapes and material representations to satisfy any given visual goal, such as providing informative visualizations to medical researchers. In this paper, we presented such a formal framework for obtaining hyper-realistic visual illusions. Quantitative and qualitative insights based on this framework can be extremely useful for the construction of targeted visual goals. Our framework emulates the experience of artists. It goes beyond photography, but we still base our framework on the limitations of humans in capturing the 3D world. Our observation is that artists despite this limitation create impressive work. Therefore,  we want to leverage this limitation by representing shapes and material parameters in a very liberal way to obtain acceptable hyper-realist results.

\begin{figure}[htbp!]
    \centering
        \includegraphics[width=0.6\textwidth]{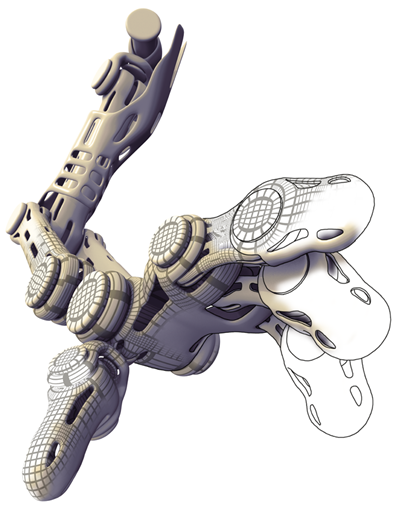}
     \caption{Augmenting the rendering of a complicated object using a mix use of styles and removing some details \cite{stanley2005digital}. }
    \label{fig_augment/stanley}
    \end{figure}

\section{Future Work}

The main advantage of our approach comes from the fact that we do not need to have very precise representations to obtain a wide variety of styles. Since our approach is not physically based\footnote{except computation of reflections during illumination stage, which we can completely ignore for shape and material estimation.}, we can view the creation of shapes and materials as an inverse problem using only statistical or mathematical methods \cite{Ozturk2006EGUK, kurt2010anisotropic, kurt2009survey, bilgili2011general, ward2014reducing, kurt2008representing, Ozturk2010CGF, Szecsi2010, bilgili2012SCCG, Ergun2012SCCG, Kurt2013TPCG, Gungor2014SIU, Kurt2017MAM, Tongbuasirilai2017ICCVW, kurt2018DEU, kurt2019DEU, Kurt2020MAM, tongbuasirilai2020TVC, kurt2021TVC, mir2022DEU, Gok2023SIU}.  

Having algebraic frameworks also simplifies the inverse problems we face for emulations and augmentations. For example, similar to Guernera et al.~\cite{guarneraetal20}, we can estimate the material parameters of figurative paintings. This estimation of material parameters can be done by computing method difference images between reference figurative paintings and rendered images. By optimizing material parameters, we try to minimize the Peak signal-to-noise ratio (PSNR) value, which is computed between reference figurative paintings and rendered images.  

\subsection{Hyper-realist Emulations}

The main application of hyper-realist emulations is to turn figurative paintings into dynamic structures. We are planning to prepare a paper on obtaining dynamic figurative paintings since figurative paintings are the motivation for hyper-realist renderings. Painters create figurative paintings mainly based on visual observations of reality. Their methods do not necessarily come from physical reality. It is important to note that long before understanding the physical rules, artists were able to capture the illusions of reality. Understanding and representing a wide range of figurative paintings can help develop practical implementations of the approach presented in this paper. 

Hyper-realist emulations can also be used to emulate photographs. These emulations can be useful to create virtual versions of reality in which we can create alternate realities by changing the light positions. Anamorphic bas-reliefs can be acceptable beyond vantage points. However, they are still acceptable from a limited point of view. If we need to have significant displacement from a single vantage point, it is good to have multiple vantage points and interpolate between them. Since the shapes are essentially 2D, it can be possible to use 2D interpolation \cite{sederberg1992physically}. 

\subsection{Hyper-Realist Augmentations}

Figurative paintings can also be augmented using the presented method. In this case, we start by emulating a figurative painting by creating proxy geometry and proxy materials. Then, we can add any new object to the scene using the same framework \cite{kiker2010using,stanley2005digital}. Figures~\ref{fig_augment//davison13} and~\ref{fig_augment/afternoon1} show two examples of hyper-realist augmentations of figurative paintings. 

Another application of hyper-realistic augmentations is the creation of medical or scientific illustrations. In such illustrations, the main goal is to provide easily understandable visual stories through images that look realistic. Scientists and doctors often realize, for example, that photographs or realistic illustrations do not tell the story. These photographs and/or illustrations need to be turned into storied visuals through hyper-realist augmentation. The main problem in these applications is to remove unnecessary details to control gaze as shown in Figure~\ref{fig_augment/stanley} \cite{stanley2005digital}. Note that our approach can also work with complete 3D data. Therefore, for medical or scientific illustrations, we use existing shape data in form of 3D models, magnetic resonance imaging (MRI) or X-ray form. 

An additional application of hyper-realistic augmentations is real-time augmentations in historical sites by adding virtual shapes to replace missing objects such as sculptures, columns, and buildings. Using sunlight directions, it is possible to add virtual shadows of virtual objects that match existing ones. In such applications, hyper-realism will also be essential. We will also present a paper discussing the details of this hyper-realist augmentation approach. 

\subsection{Validations}

This formal framework can be validated with rigorous user studies and careful statistical analysis. This paper provided the theoretical framework with some examples of proof-of-idea. We will continue to build the algorithmic foundations for this theoretical framework.  We are not planning to conduct user studies ourselves, but we are interested in supporting other researchers in evaluating these algorithms through controlled studies of human subjects using static, dynamic, and interactive visualizations. We expect that there will be static and dynamic experiments that employ side-by-side comparisons of images and animations that are created using physically accurate versus our proposed simplified algorithmic processes. Based on our methods, it is also possible to develop interactive experiments to test the importance of global illumination for task completion. Fortunately, current ray-tracing-based technologies can support some simple global illumination effects, such as shadows, simple reflection, and refraction, in real-time virtual reality settings. The availability of simple effects in real-time virtual reality (VR) is an opportunity to design interactive experiments to study the perceptual effects of simple global illumination. However, we expect these simple effects to be still out of reach for AR applications. Our findings will be useful in the future to obtain interactive AR in the future. It will also be useful to improve the performance of virtual reality applications. Moreover, more complex effects, such as radiosity or caustics, are beyond the reach of even VR applications. We will study the perceptual effects of more complex global illumination effects only through images and animations. 

\bibliographystyle{unsrtnat}
\bibliography{references}

\end{document}